\documentclass{ws-procs9x6}
\usepackage{epsfig}
\usepackage{graphicx}% Include figure files
\usepackage{bm}% bold math
\begin{document}
% Useful Definitions
\newcommand{ \snn }{${\rm \sqrt{s_{_{NN}}}}$ =} \newcommand{ \sn
}{${\rm \sqrt{s_{_{NN}}}}$} \def\Np{${\rm N_{part} }$}
\def\Npavg{${\rm \langle N_{part} \rangle}$} \def\ep{${\rm
\varepsilon_{part} }$} \def\es{${\rm \varepsilon_{std}}$}
\def\epq{${\rm \sqrt{\langle \varepsilon_{part}^{2}} \rangle}$}
\def\epr{${\rm v_{2}/\langle \varepsilon_{\rm part} \rangle }$}
\def\lsim{\mathrel{\rlap{\lower4pt\hbox{\hskip1pt$\sim$}}
\raise1pt\hbox{$<$}}} %less than or approx. symbol
\def\gsim{\mathrel{\rlap{\lower4pt\hbox{\hskip1pt$\sim$}}
\raise1pt\hbox{$>$}}} %greater than or approx. symbol 
\title{Elliptic Flow, Initial Eccentricity and Elliptic Flow fluctuations in Heavy Ion
Collisions at RHIC} \author{Rachid Nouicer \lowercase{ for the} PHOBOS
C\lowercase{ollaboration}} \address{Brookhaven National Laboratory,
Physics Department\\ E-mail: rachid.nouicer@bnl.gov}
\author{B.Alver$^4$,B.B.Back$^1$,M.D.Baker$^2$,M.Ballintijn$^4$,D.S.Barton$^2$,R.R.Betts$^6$,
A.A.Bickley$^7$,R.Bindel$^7$,W.Busza$^4$,A.Carroll$^2$,Z.Chai$^2$,M.P.Decowski$^4$,E.Garc\'{\i}a$^6$,
T.Gburek$^3$,N.George$^2$,K.Gulbrandsen$^4$,C.Halliwell$^6$,J.Hamblen$^8$,M.Hauer$^2$,
C.Henderson$^4$,D.J.Hofman$^6$,R.S.Hollis$^6$,B.Holzman$^2$,A.Iordanova$^6$,J.L.Kane$^4$,
N.Khan$^8$,P.Kulinich$^4$,C.M.Kuo$^5$,W.Li$^4$,W.T.Lin$^5$,C.Loizides$^4$,S.Manly$^8$,A.C.Mignerey$^7$,
R.Nouicer$^{2,6}$,A.Olszewski$^3$,R.Pak$^2$,C.Reed$^4$,C.Roland$^4$,G.Roland$^4$,J.Sagerer$^6$,
H.Seals$^2$,I.Sedykh$^2$,C.E.Smith$^6$,M.A.Stankiewicz$^2$,P.Steinberg$^2$,G.S.F.Stephans$^4$,
A.Sukhanov$^2$,M.B.Tonjes$^7$,A.Trzupek$^3$,C.Vale$^4$,G.J.van~Nieuwenhuizen$^4$,
S.S.Vaurynovich$^4$,R.Verdier$^4$,G.I.Veres$^4$,P.Walters$^8$,E.Wenger$^4$,F.L.H.Wolfs$^8$,
B.Wosiek$^3$,K.Wo\'{z}niak$^3$,B.Wys\l ouch$^4$} \address{\small
$^1$~Argonne National Laboratory, Argonne, IL 60439-4843, USA\\
$^2$~Brookhaven National Laboratory, Upton, NY 11973-5000, USA\\
$^3$~Institute of Nuclear Physics, Krak\'{o}w, Poland\\
$^4$~Massachusetts Institute of Technology, Cambridge, MA 02139-4307,
USA\\ $^5$~National Central University, Chung-Li, Taiwan\\
$^6$~University of Illinois at Chicago, Chicago, IL 60607-7059, USA\\
$^7$~University of Maryland, College Park, MD 20742, USA\\
$^8$~University of Rochester, Rochester, NY 14627, USA} \maketitle
\abstracts{We present measurements of elliptic flow and event-by-event
fluctuations established by the PHOBOS experiment. Elliptic flow
scaled by participant eccentricity is found to be similar for both
systems when collisions with the same number of participants or the
same particle area density are compared. The agreement of elliptic
flow between Au+Au and Cu+Cu collisions provides evidence that the
matter is created in the initial stage of relativistic heavy ion
collisions with transverse granularity similar to that of the
participant nucleons. The event-by-event fluctuation results reveal
that the initial collision geometry is translated into the final state
azimuthal particle distribution, leading to an event-by-event
proportionality between the observed elliptic flow and initial
eccentricity.}
%%%%%%%%%%%%%%%%%%%%%%%%%%%%%%%%%%%%%%%%%%%%%%%%%%%%%%%%%%%%%%%%%
\begin{figure}[!t]
\begin{tabular}{cc}
\begin{minipage}{3in}
\hspace*{-0.5cm}
\includegraphics[width=0.7\textwidth]{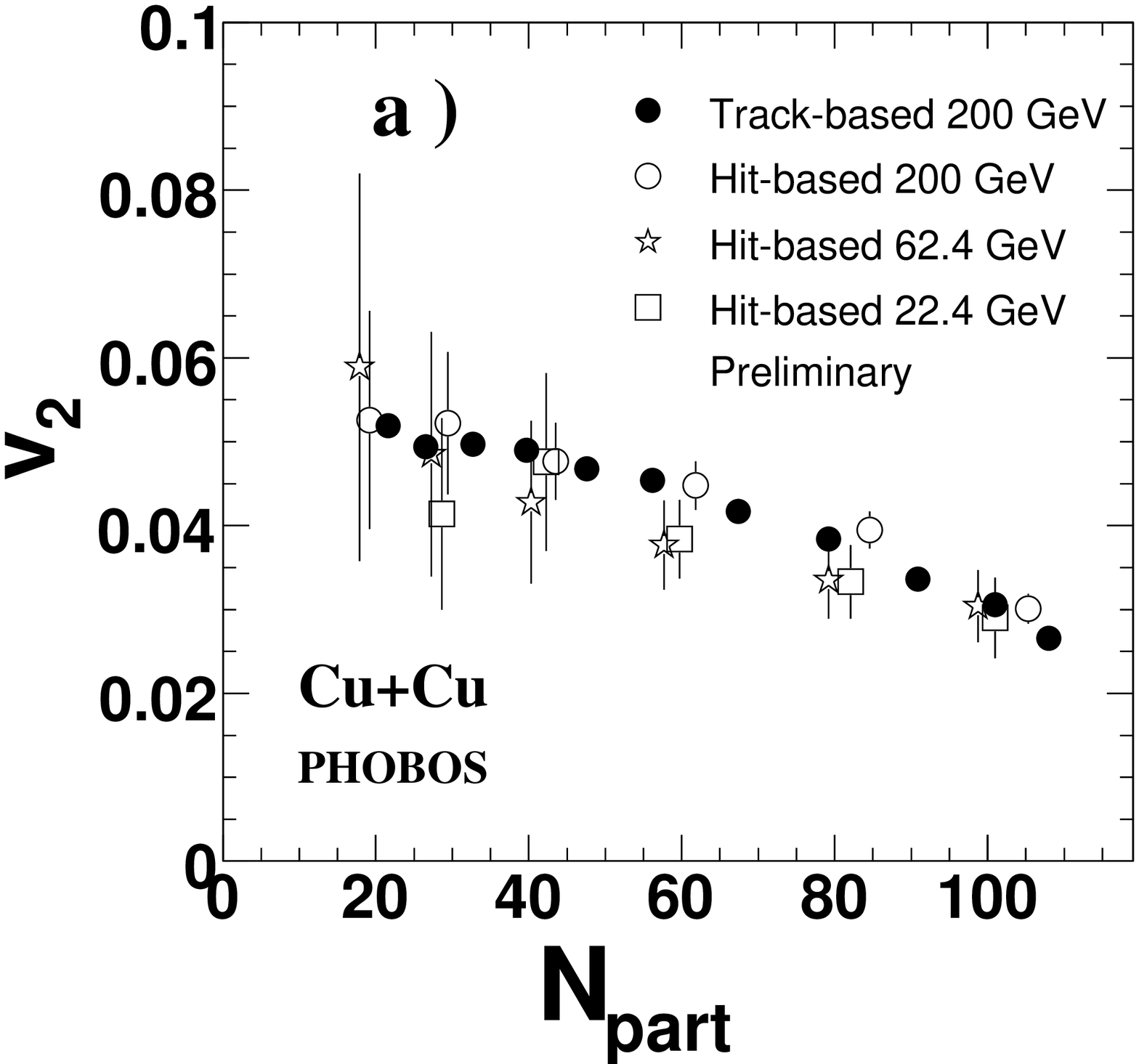}
\end{minipage}&
\hspace*{-1cm}
\begin{minipage}{3in}
\hskip -2cm 
\includegraphics[width=0.8\textwidth]{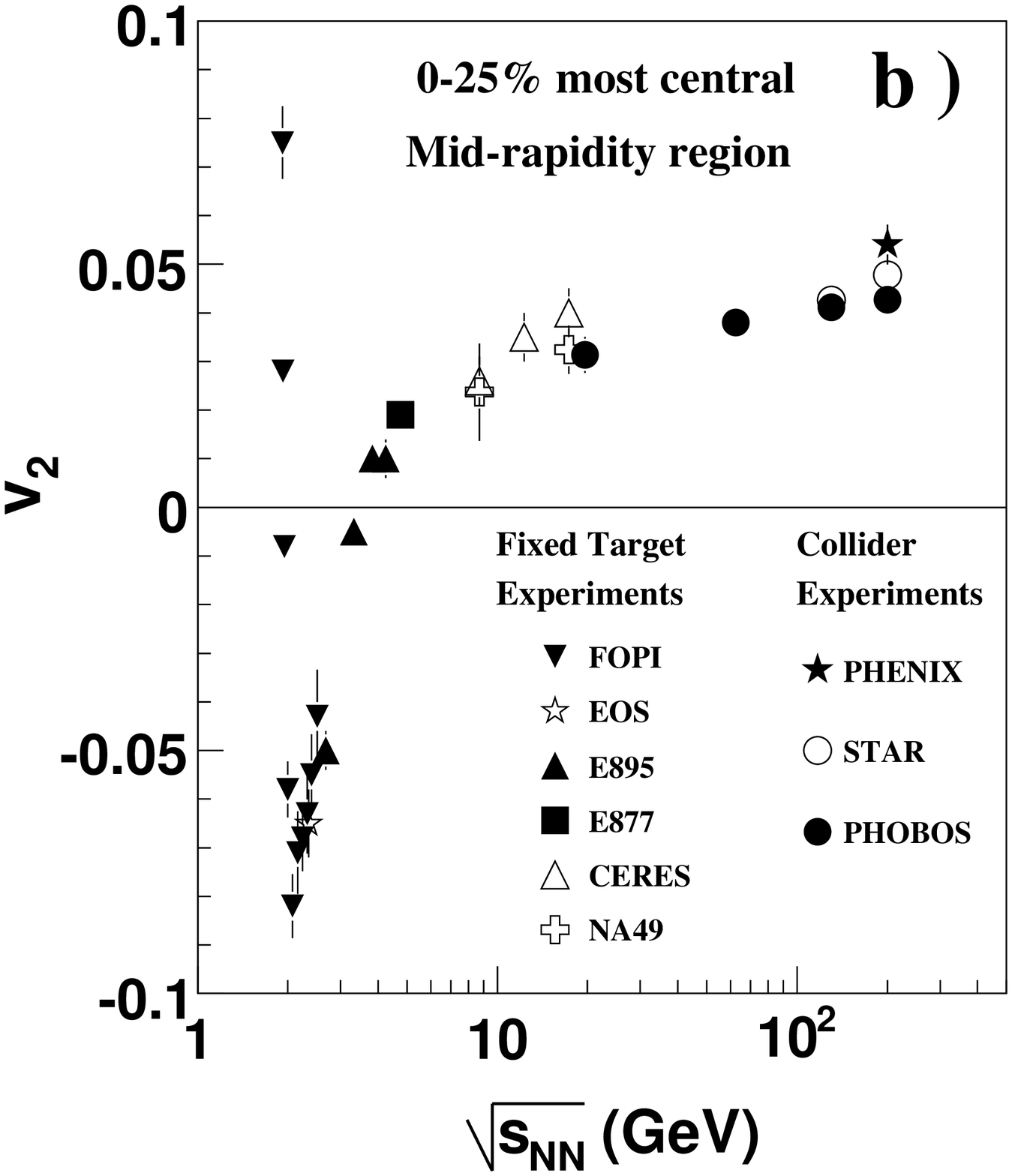}
\end{minipage}
 \end{tabular}
 \caption{ Panel a) v$_{2}$ versus N$_{\text{part}}$ for Au+Au and
Cu+Cu collisions at $\sqrt{s_{_{\it NN}}} =$ 22.4, 62.4 and 200 GeV.
Panel b) the dependence of v$_{2}$ on beam energy for mid-central
Au+Au collisions. We add to this plot PHOBOS published data for Au+Au
at $\sqrt{s_{_{\it NN}}}$ = 19.6, 62.4, 130 and 200 GeV. The bars in
the plots represent statistical errors.\label{fig:fig1}}
\end{figure}
%%%%%%%%%%%%%%%%%%%%%%%%%%%%%%%%%%%%%%%%%%%%%%%%%%%%%%%%%%%%%%%%%
\section{Introduction}
Elliptic flow has been studied extensively in nucleus-nucleus
collisions at SPS and RHIC as a function of pseudorapidity,
centrality, transverse momentum and center-of-mass
energy~\cite{SPS,RHIC}. One of the most striking observations at RHIC
is a strong event anisotropy in non-central collisions~\cite{STAR},
which is generated through the elliptically deformed overlap region of
the colliding nuclei, resulting in an eccentric distribution of matter
and anisotropic pressure gradients in the early stages of the
expansion~\cite{Olli}. From a microscopic point of view this strong
collective anisotropy is best described under the assumption of
extremely strong rescattering~\cite{Moll}, strong enough in fact to
reach the limit of continuum dynamics. To achieve such a strong
conversion of anisotropies from coordinate to momentum space,
rescattering has to be strong at very early times and local
thermalization has to occur while the geometric deformation of the
source is still large~\cite{kolb}. In this work, the comparison of the
data from Cu+Cu and Au+Au collisions measured by PHOBOS experiment at
RHIC provides new information on the interplay between initial
collision geometry (initial eccentricity) and initial particle density
in determining the observed final state flow pattern. Studies from
PHOBOS have pointed out the importance of fluctuations in the
initial-state geometry for understanding the large Cu+Cu v2 value.
%%%%%%%%%%%%%%%%%%%%%%%%%%%%%%%%%%%%%%%%%%%%%%%%%%%%%%%%%%%%%%%%%
\section{Elliptic Flow}
PHOBOS has measured elliptic flow as a function of pseudorapidity,
centrality, transverse momentum, center-of-mass
energy~\cite{PhobosFlowPRL1,PhobosFlowPRL2,PhobosFlowPRC} and,
recently, nuclear species~\cite{PhobosFlowPRL3}. In particular, the
measurements of elliptic flow as a function of centrality provide
information on how the azimuthal anisotropy of the initial collision
region drives the azimuthal anisotropy in particle production. In
figure~\ref{fig:fig1}a, we show the centrality dependence of v$_2$ at
midrapidity~($|\eta|<1$) for Cu+Cu at \snn\ 22.4, 62.4 and 200 GeV
collision energies, as obtained from our hit-based and track-based
analysis methods~\cite{PhobosFlowPRL2,PhobosFlowPRL3}. A substantial
flow signal is measured in Cu+Cu at three energies, even for the most
central events. The strength of Cu+Cu v$_{2}$ signal is surprising in
light of expectations that the smaller system size would result in a
much smaller flow signal~\cite{Chen}.  Figure.~\ref{fig:fig1}b shows
the dependence of v$_{2}$ on beam energy~\cite{Urs}. We add to this
plot PHOBOS published data for Au+Au at $\sqrt{s_{_{\it NN}}}$ = 19.6,
62.4, 130 and 200 GeV~\cite{PhobosFlowPRL2}. At low fixed target
energies (\sn $\sim$ 3 GeV), particle production is enhanced in the
direction orthogonal to the reaction plane, and v$_{2}$ is
negative. This is due to the effect that the spectator parts of the
nuclei block the matter in the direction of the reaction plane. At
higher center of mass energies, these spectator components move away
sufficiently quickly, and therefore particle production is enhanced in
the reaction plane, leading to v$_{2} > 0$. This phenomenon is
expected in hydrodynamic scenarios in which the large pressure
gradients within the reaction plane drive a stronger
expansion. However, the most important observation is that, up to the
highest center of mass energies at RHIC, the observed asymmetry
v$_{2}$ continues to grow. It would be interesting to see whether this
tendency is confirmed by the LHC data.
%%%%%%%%%%%%%%%%%%%%%%%%%%%%%%%%%%%%%%%%%%%%%%%%%%%%%%%%%%%%%%%%
\begin{figure}[!t]
\centering
  \begin{tabular}{cc}
    \begin{minipage}{3in}
  \hspace*{-1.cm}
\includegraphics[height=.23\textheight]
  {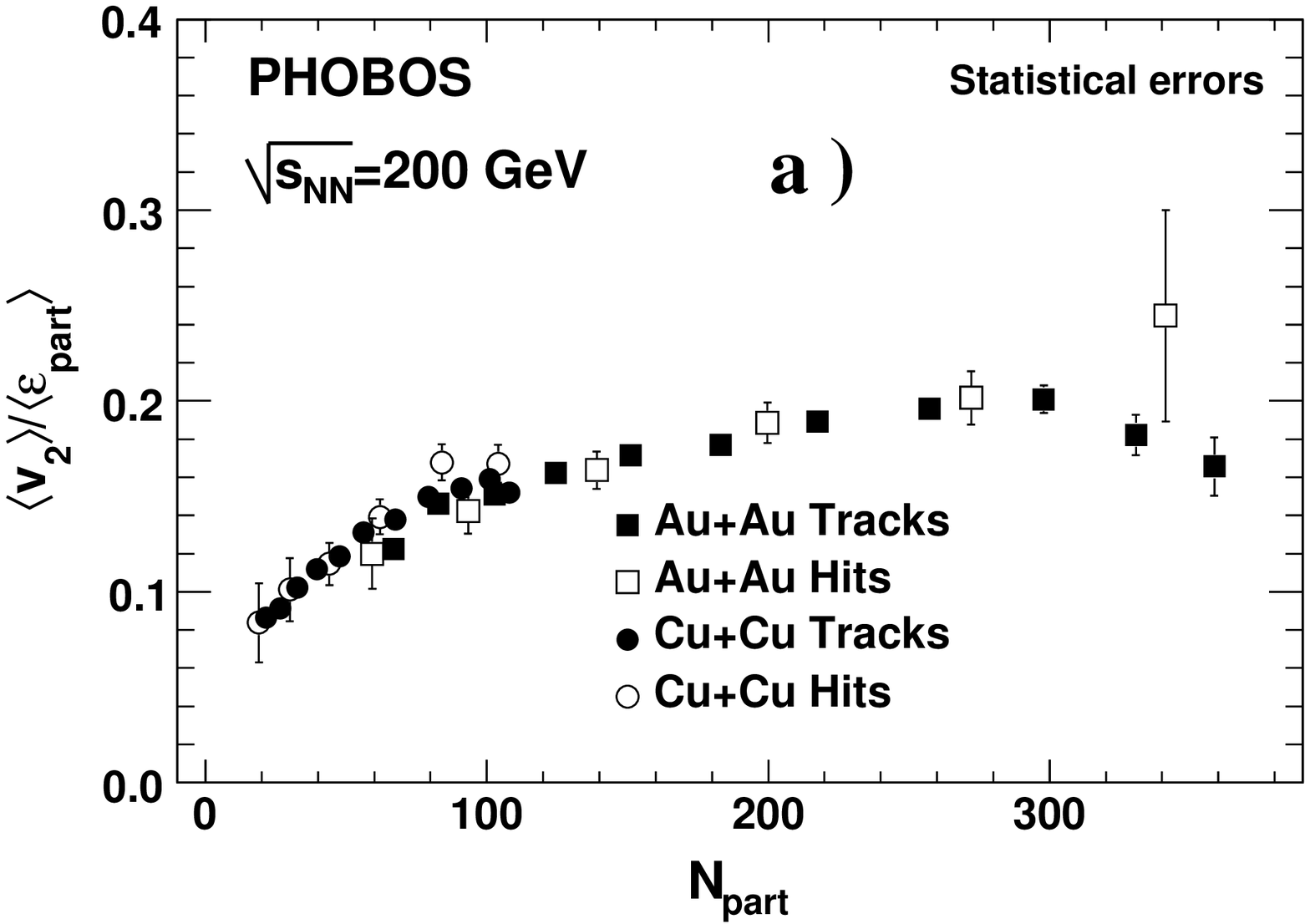}
    \end{minipage}
&
    \begin{minipage}{3in}
      \hspace*{-2.5cm}
  \includegraphics[height=.24\textheight]
	    {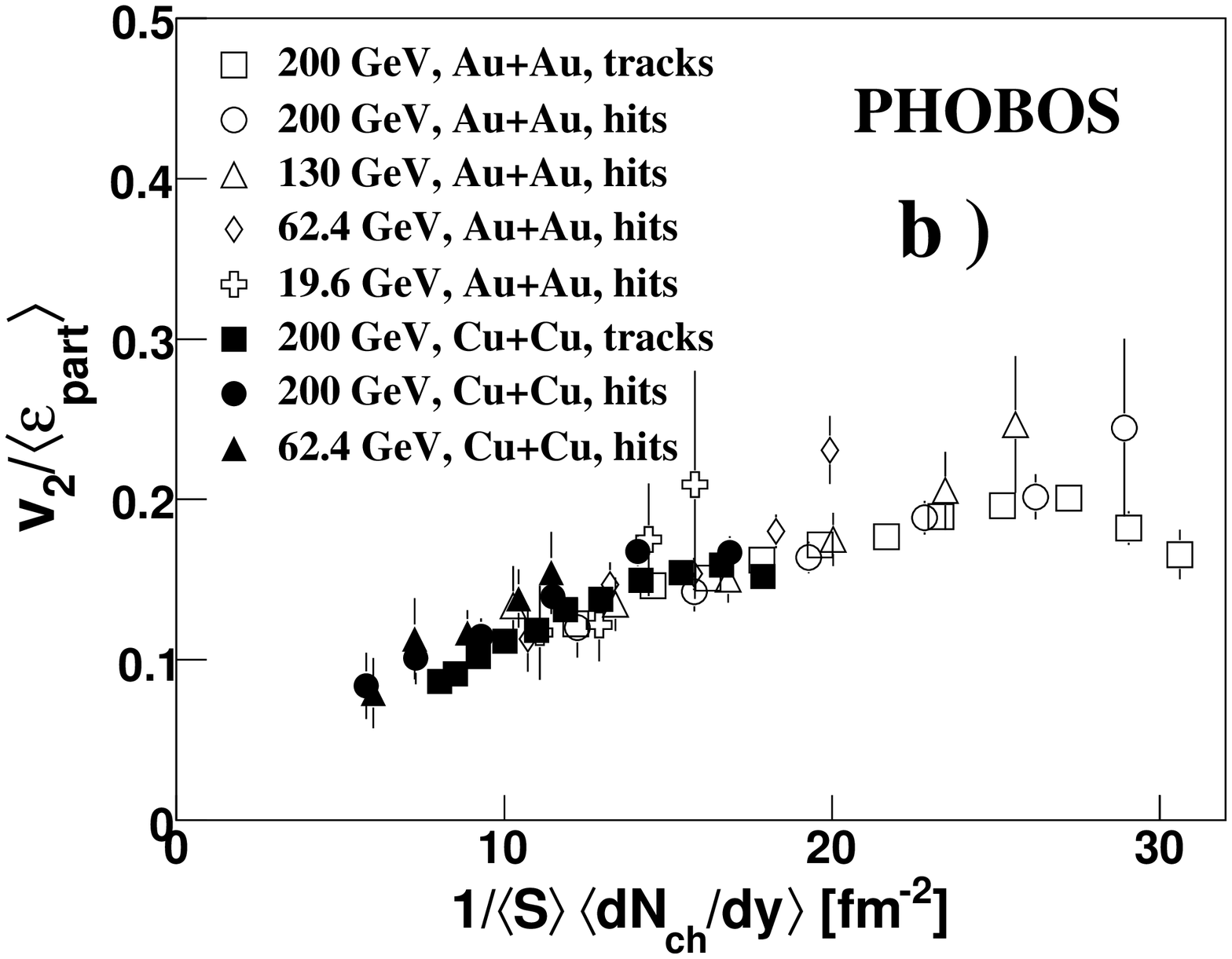}
    \end{minipage}
  \\
 \begin{minipage}{3in}
  \hspace*{-1.3cm}
\includegraphics[height=.23\textheight]
	   {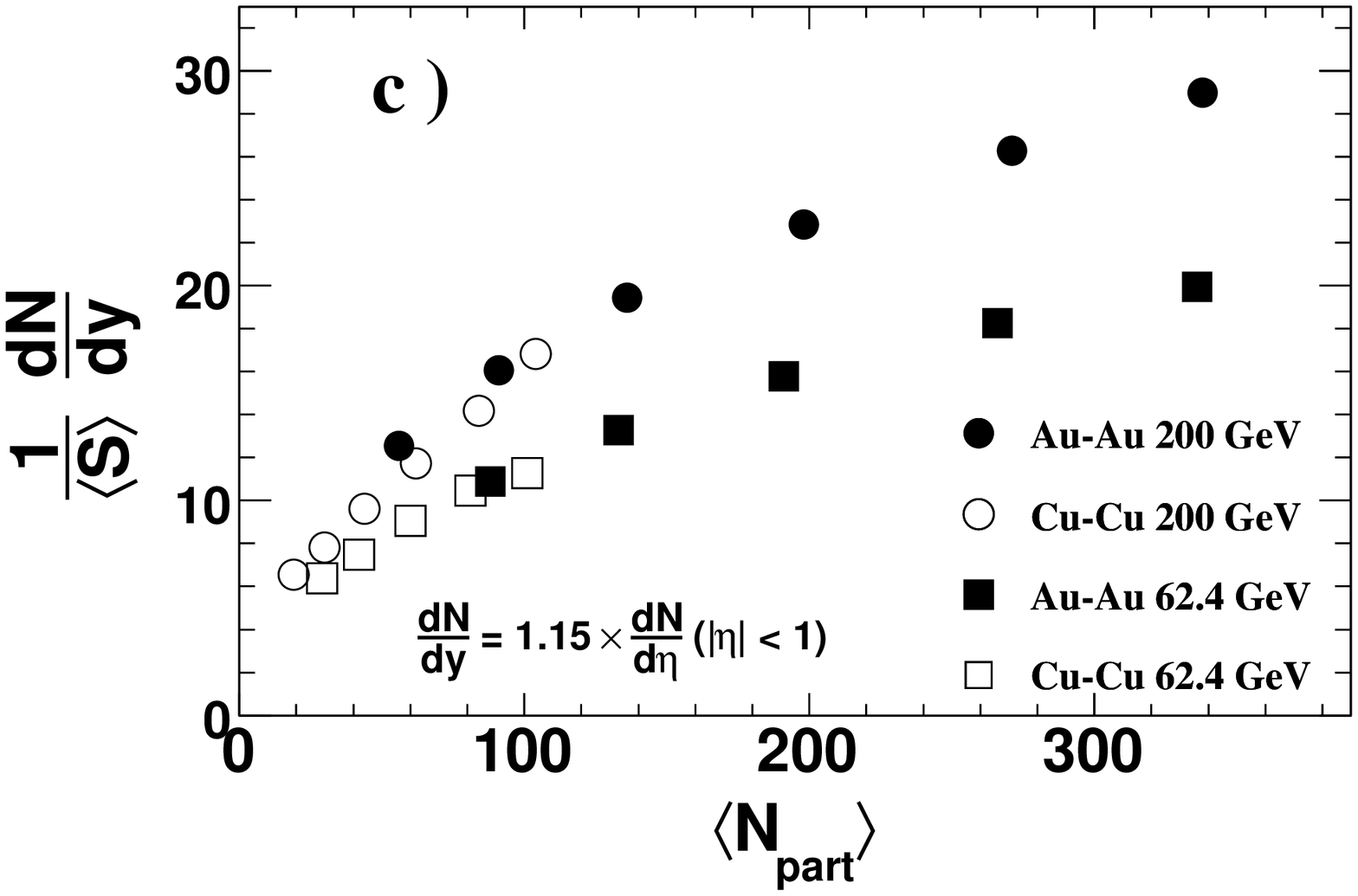}
 \end{minipage} &
\hskip -6.6cm
\begin{minipage}{2.5in} 
\hskip -0.6cm
\caption{ Panel a) \epr\ versus $N_{\text{part}}$ for Au+Au and Cu+Cu
collisions at $\sqrt{s_{_{\it NN}}} =$ 200 GeV.  Panel b) shows \epr\
as function of mid-rapidity (${\rm |\eta| < 1}$) particle area density
${\rm 1/\langle S \rangle \langle dN/dy \rangle}$ for Cu+Cu and Au+Au
collisions. The bars in the plots represent the statistical errors.
Panel c) correlation plot of midrapidity particle area density ${\rm
1/\langle S \rangle \langle dN/dy \rangle}$ as function of
$N_{\text{part}}$ calculated in Glauber MC for Au+Au and Cu+Cu
collisions at $\sqrt{s_{_{\it NN}}} =$ 62.4 and 200 GeV.  }
\end{minipage} 
\end{tabular}
\label{fig:fig2}
\end{figure}
%%%%%%%%%%%%%%%%%%%%%%%%%%%%%%%%%%%%%%%%%%%%%%%%%%%%%%%%%%%%%%%%%
\section{Initial Eccentricity}
In order to distinguish collision dynamics from purely
geometrical effects, it has been suggested that the measured v$_{2}$
should be scaled by the eccentricity of the nuclear overlap
area~\cite{Sorge}. The PHOBOS collaboration has shown that for small
systems or small transverse overlap regions, event-by-event
fluctuations in the shape of the initial collision region affect the
elliptic flow. Monte Carlo Glauber (MCG) studies have shown that the
fluctuations in the nucleon positions frequently create a situation
where the minor axis of the overlap ellipse of the participant
nucleons is not aligned with the impact parameter vector. To account
for this effect, PHOBOS has introduced the participant eccentricity
defined as~\cite{PhobosFlowPRL3}: ${\rm \varepsilon_{\rm part} =
\frac{\sqrt{(\sigma_{y}^2-\sigma_{x}^2)^2+4\sigma_{xy}^2}}{\sigma_{x}^2+\sigma_{y}^2},
 \label{eqeccpart}}$
where $\sigma_{xy}=\langle xy\rangle - \langle x\rangle\langle
y\rangle$ is the covariance.  This definition accounts for the nucleon
fluctuations by quantifying the eccentricity event-by-event with
respect to the overlap region of the participant nucleons. For
comparison of Au+Au and Cu+Cu collisions at several energies,
figure~\ref{fig:fig2}a and \ref{fig:fig2}b show \epr\ as a function of
$N_{\text{part}}$ and midrapidity particle area density, ${\rm
1/\langle S \rangle \langle dN/dy \rangle}$, respectively.
Figure~\ref{fig:fig2}c shows that at given energy, 62.4 or 200 GeV,
the Au+Au and Cu+Cu collisions selected for the same value of
$N_{\text{part}}$ leads to the same value of midrapidity particle area
density ${\rm 1/\langle S \rangle \langle dN/dy \rangle}$ calculated
in MCG. We observe in figure.~\ref{fig:fig2}a (\ref{fig:fig2}b) that
the v$_{2}$ scaled by \ep\ are similar for both Cu+Cu and Au+Au
collisions at the same value of $N_{\text{part}}$ (${\rm 1/\langle S
\rangle \langle dN/dy \rangle}$ ). It should be noted that in
figure~\ref{fig:fig2}b which has been introduced previously in
Ref.~\cite{Vol}, in the y-axis the v$_{2}(\eta)$ has been converted to
v$_{2}(y)$ by scaling the data by factor 0.9 and also in the x-axis
the dN/dy = 1.15 dN/d$\eta$ at midrapidity region, $|\eta| < 1$. This
similarity between Cu+Cu and Au+Au collisions is also observed as a
function of transverse momentum as well as in a wide pseudorapidity
range \cite{Nouicer}.
\section{Elliptic Flow Fluctuations  }
%%%%%%%%%%%%%%%%%%%%%%%%%%%%%%%%%%%%%%%%%%%%%%%%%%%%%%%%%%%%%%%%%
%########################################################################
\begin{figure}[!t]
\begin{tabular}{cc}
\begin{minipage}{3in}
\hspace*{-0.5cm}\includegraphics[width=0.75\textwidth]{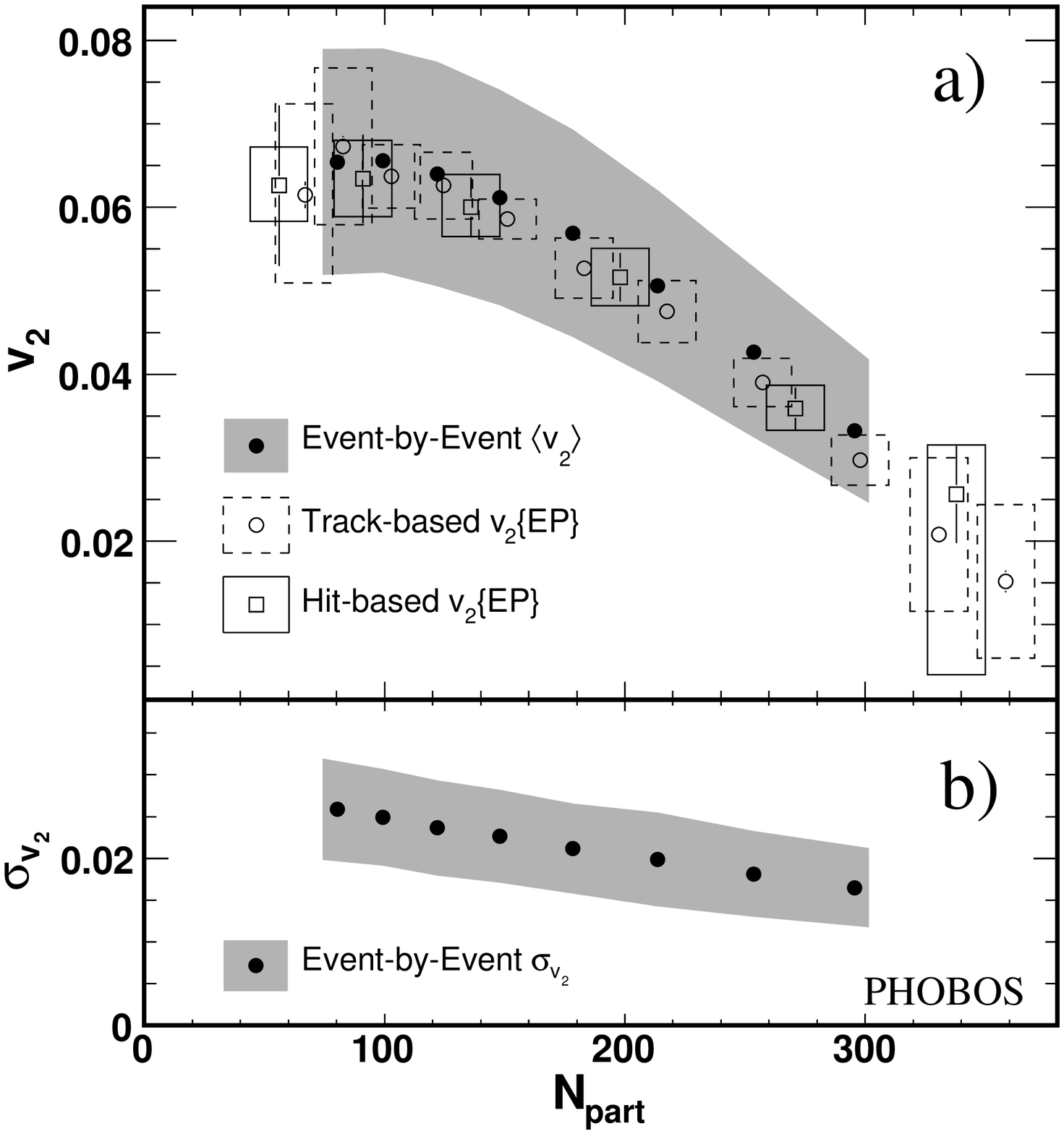}
\end{minipage}&
\hspace*{-1cm}
\begin{minipage}{3in}
\hskip -1.5cm 
\includegraphics[width=0.75\textwidth]{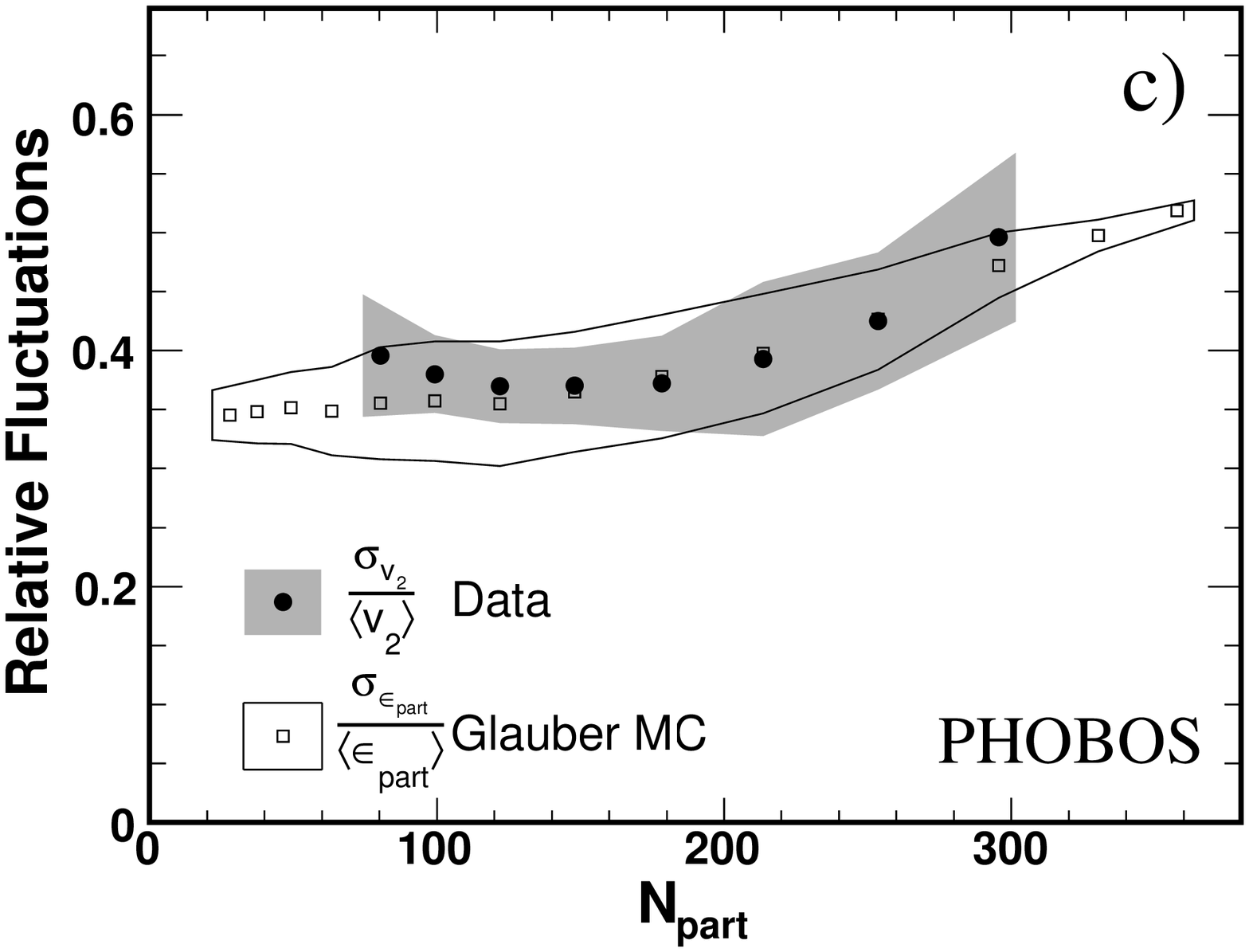}
\end{minipage}
 \end{tabular}
 \caption{ Panel a) $\langle v_2 \rangle$ and panel b) $\sigma_{v_2}$
    versus $N_{\text{part}}$ for Au+Au collisions at $\sqrt{s_{_{\it
    NN}}} =$ 200~GeV. Boxes and gray bands show 90\% C.L. systematic
    errors and the error bars represent 1-$\sigma$ statistical errors.
    The results are for $0<\eta<1$ for the track-based method and
    $|\eta|<1$ for hit-based and event-by-event methods. Panel c)
    $\sigma_{v_2}/\langle v_2 \rangle$ versus $N_{\text{part}}$ for
    Au+Au collisions at $\sqrt{s_{_{\it NN}}} =$ 200~GeV. Open squares
    show $\sigma_{\epsilon_{\text{part}}}/\langle
    \epsilon_{\text{part}} \rangle$ calculated in a Glauber MC. The
    bands show 90\% C.L. systematics errors.}\label{fig:fig3}
\end{figure}
The apparent relevance of the participant eccentricity model in
unifying the average elliptic flow results for Cu+Cu and Au+Au
collisions leads naturally to consideration of the dynamical
fluctuations of both the participant eccentricity itself as well as in
the elliptic flow signal from data. Simulations of the expected
dynamical fluctuations in participant eccentricity as a function of
N$_{\rm part}$ were performed using the PHOBOS Monte Carlo Glauber
based participant eccentricity model. Figure.~\ref{fig:fig3}a shows
the mean, $\langle \text{v}_2 \rangle$, and the standard deviation,
$\sigma_{\text{v}_2}$, of the elliptic flow parameter v$_2$ at
midrapidity as a function of the number of participating nucleons, in
Au+Au collisions at $\sqrt{s_{_{\it NN}}} =$ 200~GeV for 6--45\% most
central events \cite{PhobosFlu}. The results for $\langle
\text{v}_2 \rangle$ are in agreement with the previous PHOBOS v$_2$
measurements~\cite{PhobosFlowPRL2}, which were obtained with the
event-plane method for charged hadrons within \mbox{$|\eta| \! < \!
1$}.  The uncertainties in $dN/d\eta$ and $\text{v}_2(\eta)$, as well
as differences between HIJING and the data in these quantities,
introduce a large uncertainty in the overall scale in the
event-by-event analysis due to the averaging procedure over the wide
pseudorapidity range. The event-plane method used in the previous
PHOBOS measurements has been proposed to be sensitive to the second
moment, $\sqrt{\langle \text{v}_2^2 \rangle}$, of elliptic
flow~\cite{OllitraultPri}. The fluctuations presented in this work
would lead to approximately 10\% difference between the mean, $\langle
\text{v}_2 \rangle$, and the RMS, $\sqrt{\langle \text{v}_2^2
\rangle}$, of elliptic flow at a fixed value of
$N_{\text{part}}$. Most of the scale errors cancel in the ratio,
$\sigma_{\text{v}_2}/\langle \text{v}_2 \rangle$, which defines
``relative flow fluctuations'', shown in figure.~\ref{fig:fig3}b as a
function of the number of participating nucleons \cite{PhobosFlu}. We observe large
relative fluctuations of approximately 40\%. MC studies show that the
contribution of non-flow correlations to the observed elliptic flow
fluctuations is less than 2\%. Figure~\ref{fig:fig3}c shows
$\sigma_{\epsilon_{\text{part}}}/\langle \epsilon_{\text{part}}
\rangle$ at fixed values of $N_{part}$ obtained in a MC Glauber
simulation.  The 90\% confidence level systematic errors are estimated
by varying Glauber parameters as discussed in
Ref.~\cite{PhobosFlowPRL3}.  A striking agreement between the relative
fluctuations in the Glauber model participant eccentricity predictions
and the observed elliptic flow fluctuations is seen over the full
centrality range under study.  The observed agreement suggests that
the fluctuations of elliptic flow primarily reflect fluctuations in
the initial state geometry and are not affected strongly by the later
stages of the collision.
%%%%%%%%%%%%%%%%%%%%%%%%%%%%%%%%%%%%%%%%%%%%%%%%%%%%%%%%%%%%%%%%
\section{Summary }
We have performed a comprehensive examination of the elliptic flow of
charged hadrons produced in Cu+Cu and Au+Au collisions at \snn\ 19.6,
22.4, 64.4 and 200 GeV. We also presented the measurements of
event-by-event fluctuations for Au+Au collisions at \snn\ 200 GeV. The
comparison of the data from Cu+Cu and Au+Au collisions provides new
information illustrating that the participant eccentricity is the
relevant geometric quantity for generating the azimuthal asymmetry
leading to the observed flow. The magnitude of event-by-event
fluctuations agree with predictions for fluctuations of the initial
shape of the collision region based on the Glauber model. These
results provide qualitatively new information on the initial
conditions of heavy ion collisions and the subsequent collective
expansion of the system.

\newcommand{\etal} {$\mathrm{\it et\ al.}$}

\end{document}